# Structure Formation with Decaying Neutrinos

M. White*, G. Gelmini** and J. Silk*

*Center for Particle Astrophysics and Departments of Astronomy and Physics
University of California, Berkeley, CA 94720-7304

**Department of Physics, University of California
Los Angeles, CA 90024-1547

**Abstract**

We consider the effects of a massive, unstable neutrino on the evolution of large–scale structure and anisotropies in the cosmic microwave background. Comparison with large–scale structure data allows us to rule out a wide range of masses and lifetimes for such neutrinos. We also define a range of masses and lifetimes which delay matter–radiation equality and improve the agreement with the data of Cold Dark Matter models with critical density.

PACs Numbers: 98.70Vc, 04.30.+x, 12.10.Dm, 98.80.Cq

## 1. Introduction

Large-scale structure is facing a crisis in the post-COBE [1,2,3] era, in that the normalization to cosmic microwave background fluctuations on large angular scales is strongly constraining the choice of allowable matter power spectra, $P(k)$. In the context of inflationary models, with $\Omega = 1$ in mostly Cold Dark Matter (CDM) and primordial scale-invariant fluctuations, $P(k)$ is fully specified by a "shape" and normalization. The latter is fixed by the amplitude of the temperature fluctuations in the Cosmic Microwave Background radiation (CMB) measured by COBE [4]. For models normalized to COBE one finds both the amplitude at small scales and the "shape" of the power spectrum predicted by CDM are inconsistent with observations of large-scale structure (LSS).

On small scales the normalization of the power spectrum is usually expressed as $\sigma_8$, the ratio of the r.m.s. mass fluctuations to the galaxy number fluctuations, both averaged over randomly located spheres of radius $8h^{-1}$Mpc. Observationally the latter has unit amplitude on this scale [5,6], so that $\sigma_8$ directly measures the mass density fluctuations. Observations of pairwise galaxy velocities (on Mpc scales) and cluster abundances, effectively probing the power spectrum near 10 Mpc, require $\sigma_8 \approx 0.5$. On slightly larger scales bulk flows measured over $\sim 10-70 h^{-1}$Mpc [7,8,9] would require $\sigma_8 \Omega^{0.6} = 1.3(\pm 0.5)$. Standard CDM, normalized to COBE, has $\sigma_8 = 1.3 \pm 0.1$ [4].

A phenomenological large-scale power spectral "shape" fitting parameter [10], which basically measures the horizon scale at matter–radiation equality, may be defined by (see Eq.(13))

$$\Gamma \approx \Omega_0 h \left(\frac{g_*}{3.36}\right)^{-1/2}, \tag{1}$$

where $g_*$ counts the relativistic degrees of freedom and $g_* = 3.36$ corresponds to the standard model with photons and three massless neutrino species. The large–scale galaxy distribution is phenomenologically fit if the "shape parameter" $\Gamma \approx 0.25 \pm 0.05$ [11], compared with $\Gamma \simeq 0.5$ for CDM (with $\Omega_0 = 1$ and $h = 0.5$). One can achieve the desired fit in low density models, with either open CDM or $\Lambda$CDM, where $g_* = 3.36$ and $\Omega_0 \simeq \Gamma/h$ (such models still have some trouble fitting the normalization of $P(k)$ however [12]). A critical density model is also possible if $h$ is decreased [13] or alternatively if $g_*$ instead is increased. One means of enhancing $g_*$, which we shall explore here, is to fine-tune the mass and decay time of one neutrino species, which we shall refer to as $\nu_\tau$.

If the neutrino is massive (1–10 MeV) and decouples non-relativistically, the decay timescale must be short ($\lesssim 100$s) to avoid excessively perturbing nucleosynthesis [14,15]. (Decaying neutrinos with short lifetimes were first proposed as a means of reconciling the normalization on galaxy scales with COBE in [16], and have received renewed attention recently [17,18].) For such short lifetimes the early epoch of matter (neutrino) domination occurs on such small scales that it is of little interest for seeding any early galaxy formation. In contrast, the case of a relativistically decoupling, massive neutrino has no effect on nucleosynthesis, and LSS and CMB provide the best constraints. Bardeen, Bond & Efstathiou [19] noted that a neutrino with mass in the keV range and a short lifetime ($10^4$yr in their example) would provide extra power on galaxy scales, and this idea was further pursued in the context of the 17keV neutrino with a lifetime of the order of years by Bond & Efstathiou [20]. In this paper we consider the effects of a relativistically decoupling $\nu_\tau$ on the evolution of density perturbations, for a range of masses and lifetimes. Such a neutrino changes both the shape of the power spectrum of CDM and the intermediate–scale angular anisotropy in the CMB. We have studied in detail the resulting radiation power spectrum for the range of neutrino masses and lifetimes which yield a LSS shape parameter in the desired range. We find that decaying neutrinos are capable of leaving a distinctive, although subtle, signature in the CMB radiation power spectrum.

## 2. Simple Calculation

The idea that neutrinos may have a mass and be unstable to decay is well motivated in many extensions of the standard model of particle physics (see section 5). Indeed if neutrinos with standard annihilation cross section have a mass such that

$$\sum m_\nu \gtrsim 92 \, (\Omega_0 h^2) \, \text{eV}, \tag{2}$$

they are required to decay so that their energy density today does not overclose the universe.

In order to build some intuition for the problem, in this section we review and extend a "back-of-the envelope" calculation of the matter power spectrum [21,22,23,24]. The interesting feature of a cosmology with decaying neutrinos is that, for some range of mass $m$ and lifetime $\tau$, the universe has *two* epochs of radiation and matter domination. At very early times the universe is radiation dominated, as in the standard cosmology, but for large $m$ there is a redshift, $z_1$, at which the neutrinos become non-relativistic and begin to dominate the energy density of the universe. The universe then enters its first matter-dominated phase, which lasts until the neutrino decays (into relativistic products) at $z_{\text{dec}}$ at which point the universe becomes radiation dominated once again. The energy in the decay products redshifts relative to that of non-relativistic matter until the universe becomes matter dominated for the second time at $z_2$ by a non-relativistic component, Cold Dark Matter ($X$) plus baryons ($B$), with present density $\Omega_{NR} = \Omega_X + \Omega_B$. Assuming the massive neutrinos were relativistic and had standard annihilation cross sections when they decoupled from the thermal bath ($T \sim 1\text{MeV}$) we estimate

$$\begin{aligned}
1 + z_1 &= 2.9 \times 10^5 (m/\text{keV}) \left[1 + 0.2(m/\text{keV})^{4/3}(\tau/\text{yr})^{2/3}\right]^{-1/3} \\
1 + z_{\text{dec}} &= 1.3 \times 10^6 (m/\text{keV})^{-1/3}(\tau/\text{yr})^{-2/3} \\
1 + z_2 &= 2.8 \times 10^4 \left(\Omega_{NR} h^2\right) \left[1 + 0.2(m/\text{keV})^{4/3}(\tau/\text{yr})^{2/3}\right]^{-1} \quad .
\end{aligned} \quad (3)$$

Here we have made the approximations of instantaneous decay at $\tau$ and sudden change of matter and radiation domination and we take the present temperature $T_0 = 2.73\text{K}$. For simplicity we assume the decay products stay relativistic until now, thus the bracket in the expression for $1 + z_1$, which corresponds to the ratio of effective entropy degrees of freedom $[g_s^*(T_1)/g_s^*(T_0)]$ is not one (as assumed in [23], for example).

The constraint on the mass and lifetime from the requirement that the tau neutrino decay products do not overclose the universe is

$$\left(\frac{m}{\text{keV}}\right)^2 \left(\frac{\tau}{\text{yr}}\right) \lesssim 1.2 \times 10^8 \left(\frac{t_0}{1.3 \times 10^{10}\text{yr}}\right) \left(\Omega_0 h^2\right)^2 \quad , \quad (4)$$

where whenever necessary we take $\Omega_0 = 1 \simeq \Omega_{NR}$, $t_0 = 1.3 \times 10^{10}\text{yr}$ and $h = 0.5$. A more restrictive limit is obtained from structure formation arguments, which we deal with next.

The power spectrum we observe today is modified from its "primordial" form by time evolution in the expanding universe. The relation between the initial and evolved power spectra is given by the transfer function, $T(k)$, (see Eq.(6,7) below). A crude approximation to $T(k)$ may be obtained by realizing that, inside the horizon, structure can only grow while the universe is in a matter dominated phase. In this case the fractional density contrast $\delta\rho/\rho \equiv \delta \propto \eta^2$, with $d\eta = dt/a$ the conformal time. Thus those modes which enter the horizon (technically the Jeans scale, but for our purposes the two are equivalent) during a radiation dominated epoch have their growth suppressed relative to modes which enter during matter domination. The modes which enter the horizon ($k = aH$) at the epochs of Eq.(3) have comoving wavenumber

$$\begin{aligned}
k_{\text{eq2}} &= 3.33 \times 10^{-4}(h\text{Mpc}^{-1})\sqrt{1 + z_2} \\
k_{\text{dec}} &= 3.33 \times 10^{-4}(h\text{Mpc}^{-1})(1 + z_{\text{dec}})/\sqrt{1 + z_2} \\
k_{\text{eq1}} &= 3.33 \times 10^{-4}(h\text{Mpc}^{-1})\sqrt{(1 + z_1)(1 + z_{\text{dec}})/(1 + z_2)} \quad .
\end{aligned} \quad (5)$$

Based on this argument we would then estimate

$$T(k) = \begin{cases} 1 & k < k_{\text{eq2}} \\ (k_{\text{eq2}}/k)^2 & k_{\text{eq2}} < k < k_{\text{dec}} \\ (k_{\text{eq2}}/k_{\text{dec}})^2 & k_{\text{dec}} < k < k_{\text{eq1}} \\ (k_{\text{eq2}} k_{\text{eq1}}/k_{\text{dec}} k)^2 & k_{\text{eq1}} < k \end{cases} \quad (6)$$

where

$$P(k) \propto k T(k)^2 \quad , \quad (7)$$

and the constant of proportionality is fixed by COBE for standard CDM to be about $6 \times 10^5 (h^{-1} \text{Mpc})^4$. If we measure the amplitude of the power spectrum by $\sigma_8$, then

$$\frac{\sigma_8}{\sigma_8(\text{CDM})} \approx \frac{T(k = 0.2\, h\text{Mpc}^{-1})}{T(k = 0.2\, h\text{Mpc}^{-1}; \text{CDM})} \tag{8}$$

For a limit, suppose we now require that $\sigma_8$ not be more than a factor of 10 different from the CDM number. This means that

$$\left(\frac{m}{\text{keV}}\right)^2 \left(\frac{\tau}{\text{yr}}\right) \lesssim 200 \qquad [\text{simple } T(k)] \tag{9}$$

Here we have assumed that the neutrinos were relativistic when they decoupled, thus this bound applies to $m < 1\,\text{MeV}$. The bound in Eq.(9) coincides with the usual limit based on structure formation arguments [22], obtained by requiring the decay products to become subdominant at recombination, i.e. $\rho_{\text{products}} < \rho_{NR}$ at $t_{\text{rec}}$. As we shall see, such simple minded power–law approximations to $T(k)$ are not terribly accurate, and we shall derive more precise limits below.

## 2. Numerical Calculation

As shown in Fig. 1 approximating the transfer function in the manner outlined above is not quantitatively reliable. This is because the concept of a sudden "horizon crossing" is not well justified. To obtain a reliable estimate of $T(k)$ requires more than the arguments presented above and we have followed and generalized the treatment outlined in [20]. Specifically we have evolved the coupled fluid, Einstein and Boltzmann equations for perturbations in a cosmological simulation with decaying neutrinos. The background space time and unperturbed energy densities were evolved until the present using

$$\begin{aligned} \rho_m' &= -(3/a)\rho_m \\ \rho_\gamma' &= -(4/a)\rho_\gamma \\ \rho_\nu' &= -(4/a)\rho_\nu + (1/\tau h)\rho_\tau \\ \rho_\tau' &= -(3/a)(\rho_\tau + P_\tau) - (1/\tau h)\rho_\tau \\ \eta' &= (ah)^{-1} \end{aligned} \tag{10}$$

where $h = \dot{a}/a$, primes denote differentiation w.r.t. the scale factor $a$, and dots w.r.t. the conformal time $\eta$. We set $a_0 = 1$ and $H_0 = 50\,\text{kms}^{-1}\text{Mpc}^{-1}$ as a compromise between direct measurements and age determinations. The evolution is started when all the neutrinos are relativistic so that

$$\begin{aligned} \rho_\nu &= 2r_\nu \rho_\gamma \\ \rho_\tau &= r_\nu \rho_\gamma \end{aligned} \tag{11}$$

with $r_\nu = (7/8)(4/11)^{4/3}$. The pressure of the massive neutrinos is given by

$$\frac{P}{\rho} = f(x) \frac{1}{3 + xK_1(x)/K_2(x)} \quad \text{with } x = \frac{kT}{mc^2} \tag{12}$$

where the expression with $f = 1$ is the result assuming Maxwell-Boltzmann statistics ($K_i$ is a modified Bessel function) and $f$ is a correction factor fit to a numerical calculation of $P/\rho$ for Fermi-Dirac statistics. The analytic expression gives the asymptotic behaviour for large and small $x$, making it easy to fit for the correction factor.

For the models in Table 1 we followed the evolution of the perturbations in the synchronous gauge (see e.g. [20,25]) treating the baryon-photon fluid as tightly coupled for all time. This causes damping of $\delta_X$ at small scales unless $\Omega_B \ll 1$, but since $T(k)$ is not very sensitive to $\Omega_B$ this limit is justified. Our treatment mirrors that of [20] except that we follow the decaying neutrinos from there relativistic to non-relativistic phases. Also the collision term in the Boltzmann equation for the relativistic decay products (which we take to be $\nu_e$ and $\nu_\mu$) is proportional to $\Delta_\tau - 2\Delta_\nu$, to be compared with $\Delta_\tau - \Delta_\nu$ in [20]. The factor 2 comes from assuming a 2-body final state to conserve momentum, and has almost no effect on the transfer functions we compute (the main effect is that it moves the second "plateau" in $T(k)$ to slightly higher $k$, but this "plateau" will not play a role in our studies). When computing the CMB anisotropies (section 4) we relax the approximation of tight coupling of the photon-baryon fluid and follow the full photon distribution function carefully including the period of recombination. We have compared our results for $m = 0$ and a range of $\Omega_B$ and $h$ with power spectra from Sugiyama [26] and Dodelson [27] and find agreement to better than 3% in power (1.5% in temperature).

### 3. Large Scale Structure

We show in Table 1 some measures of large-scale power for 16 models with $m$ and $\tau$ in an interesting range. All models have been normalized to have CMB temperature fluctuations equal to that measured by COBE. For the models which are viable, the relative normalization of the large-angle CMB fluctuations and the matter power spectrum differ by $\lesssim 5\%$ from that in the standard CDM scenario. We see that for low masses and short lifetimes the results are not very different from normal CDM, whereas for large masses and lifetimes they are inconsistent with observations. Using the simple calculations of the preceding section we can see that the decaying neutrinos dominated the energy density of the universe at decay only if $(m/\text{keV})^2(\tau/1\text{yr}) \gtrsim 10$. Thus for $\sigma_8 \gtrsim 1$ in Fig. 2 the decaying neutrinos and their decay products never dominate the energy density of the universe.

For the lighter neutrinos no effect is seen until longer lifetimes than we have considered here. We have not computed $T(k)$ for models with such long lifetimes, since in this case the models are similar to hot or mixed dark matter scenarios where neutrino free streaming causes exponential damping of power on small scales [28]. In this case one has to follow the full distribution function for the "light" neutrinos and numerically integrate over momentum for each mode, which is a computationally intensive task. Furthermore, because of the exponential damping, obtaining sufficient small scale power requires one to have more power at $\sim 100\text{Mpc}$ than LSS would indicate: the LSS data seems to be best fit by a change in the length scale at matter-radiation equality (i.e. a change in $\Gamma$), not damping of power on small scales [11]. For the masses and lifetimes we have considered the exponential damping sets in at a scales $k >$few $\text{Mpc}^{-1}$ (estimated from the horizon size when the neutrinos decay), and does not affect the arguments in this paper.

We note that for decaying neutrino models with $m \sim \text{keV}$, the second period of matter domination leads to more very small scale power than in the standard scenario. This could be useful for early structure formation models, however quantitative results are hard to obtain since the details of structure formation in the non-linear regime are not well understood. We show in Fig. 3 transfer functions for two of the models considered here, the first is a model with $\Gamma \simeq 0.25$ and the second is an extreme model from Table 1. Notice that only in the extreme case is the plateau from the first period of matter domination clearly visible, and even then we find $T(k) \sim k^{-1/2}$ rather than constant.

Efstathiou, Bond & White [12] quote a fitting function

$$T(k) = \left[1 + \left(ak + (bk)^{3/2} + (ck)^2\right)^\nu\right]^{-1/\nu} \tag{13}$$

with $a = (6.4/\Gamma)\text{Mpc}$, $b = (3.0/\Gamma)\text{Mpc}$, $c = (1.7/\Gamma)\text{Mpc}$ and $\nu = 1.13$. For a neutrino of mass $m_{10} \times 10\text{keV}$ and a lifetime $\tau$ in years they obtain

$$\Gamma \approx \frac{\Omega_0 h}{\sqrt{0.861 + 3.8(m_{10}\tau)^{2/3}}} \tag{14}$$

which is numerically in rough agreement with our results for $m \sim 10\text{keV}$. Note however that there is not a degeneracy between $m_{10}$ and $\tau$ as this formula suggests, but as shown in Fig. 2,4, it is better to use $m^2\tau$. This can be understood as a change in the energy density of radiation after the decay (or $g_*$) which moves the second period of matter-radiation equality (see Eq.(3,4)). The extra energy in relativistic products comes from the energy "stored" in the $\nu_\tau$ component while it is nonrelativistic. During this phase (which starts at $z_1$ and ends at $z_{\rm dec}$) the energy density grows compared with that of the relativistic neutrinos $\propto a \sim t^{2/3}$. Since $z_1$ is determined by $T \sim m$ and $a \sim t^{1/2}$ while the universe is radiation dominated (before $z_1$) we find the energy is enhanced by an amount depending on $(m^2\tau)^{2/3}$.

Fig. 4 shows the narrow region of masses and lifetimes that give a good fit to LSS, i.e. for which $\Gamma$ is between 0.2 and 0.3. The models discussed in the next section are within this region. Note that within the assumptions we have made these considerations can rule out a large region of $(m, \tau)$ space which are inaccessible to laboratory experiments. In the limit of larger and larger masses, where the lifetimes becomes correspondingly shorter, the details of the decay process become irrelevant for all scales of astrophysical interest. For the purposes of LSS one has merely increased the number of relativistic degrees of freedom in the model and thus delayed matter-radiation equality. One could consider this as an increase in the "effective" number of massless neutrino species. We show in Fig. 4 an extrapolation of our results to the masses and lifetimes of the $\tau$CDM scenario [14]. Note the approximate degeneracy of $\Gamma$ with $m^2$ and $\tau$ appears to hold well over the whole range.

### 4. Cosmic Microwave Background Anisotropies

From the preceding discussion we see that we can fit the LSS data with a combination of masses and lifetimes in the keV and year range. The CMB also probes density fluctuations on scales in the $10-1000$ Mpc range, albeit at a much earlier epoch: $z \sim 1000$. Thus one is led to consider the effect of decaying neutrinos on the CMB. In Fig. 5 we show the angular power spectrum of the CMB anisotropies in a model which has been adjusted to have the correct shape to fit the LSS data (i.e. $\Gamma \simeq 0.25$, [11]). One obvious feature is the increase in power on degree scales, which arises due to the smaller sound horizon at last scattering. The second thing to note is the observable shift in the secondary and tertiary peaks to higher multipole moments (the visibility function in these models is changed only at very low $z$ as expected). This result is similar to that of ref. [14] who consider a very massive neutrino which decays on a much shorter time scale. In fact, for the models that lie within the band of Fig. 4, we find that explicitly following the decaying neutrino, or artificially increasing the number of massless neutrino species to give the same $\Gamma$, lead to almost indistinguishable CMB spectra. This is because the range of scales probed by the CMB is large enough that all scales were well outside the horizon when the neutrino decayed.

With the current rate of advance of CMB anisotropy measurements we expect that a measurement of the *amplitude* of the peaks in the CMB spectrum should be possible in the next few years. If the measurements continue to prefer a lower peak height then the model in Fig. 5 would be ruled out. To salvage this model a lower $\Omega_B$, tilted initial power spectrum, inclusion of tensor modes or reionization would be required. The shift of all the peaks to higher $\ell$ is a more robust indication of delayed matter-radiation equality. We have explicitly checked that the *position* of the second and third peaks is fixed by $\Gamma$ alone in models with decaying neutrinos, artificially increased number of massless neutrinos, and lower $h$. None of the complications mentioned above (changing $\Omega_B$, etc) would change the position of the second and third peaks, however the amplitude of these peaks is dependent on all of them, which makes quantitative comparison with experiment difficult. A further technical difficulty is that if $\Omega_B$ is lowered and one wishes to maintain $\Omega_B h^2$ at the nucleosynthesis value [29,30,31], it becomes necessary to increase $h$ (which would be hard to reconcile with the age estimates) and hence find a new $m$ and $\tau$ which give the same $\Gamma$.

Decaying neutrino models have slightly more small scale power due to the presence of the first period of matter domination, hence it is not implausible to assume that early object formation and reionization could have occurred. If this were the case one would expect the anisotropy at small scales to be damped. Note that in our "preferred" models however, the amount of extra power on small scales is small, and details of structure formation are not well understood.

## 5. Particle Physics Models

Radiative decays, $\nu_\tau \to \nu\gamma$ where $\nu$ stands for a lighter neutrino ($\nu_e$ or $\nu_\mu$), in the preferred range shown in Fig. 4, are ruled out by several cosmological and astrophysical constraints (see [32,33]). It will suffice to mention one of them, based on the absence of photons arriving in conjunction with the neutrinos from supernova SN1987A. This limit requires the radiative lifetime of a $\nu_\tau$, with mass between 100 eV and 1 MeV, to be $\tau_{\rm rad} \gtrsim 2.5 \times 10^7$s [34]. This still leaves the possibility of invisible decays, e.g. $\nu_\tau \to 3\nu$ and $\nu_\tau \to \nu\phi$, where $\phi$ is a very weakly interacting boson of zero or small mass, most naturally a Goldstone boson associated with a spontaneously broken leptonic global symmetry, called Majorons in most models.

Models with acceptable decay modes of this type were produced [24,35,36] for the heavier $\nu_\tau$ of [14] and for the now dead 17keV neutrino (see for example [37] and references threin). This shows the feasibility of models with long lived neutrinos decaying into invisible particles, and in our case model building should be much simpler. What considerably complicated the models for the 17keV neutrino was its alleged large mixing with the $\nu_e$, which is not a requirement in our case. In fact the mixing angles that appear in the decay rates of the modes just mentioned are in general not the same as those that appear in the mass matrix, i.e. are not those measured in oscillation or appearance or disappearance experiments, since operators with different flavour structure can contribute. Without this constraint on the mixing from experiment, the $\nu_3-\nu_e$ mixing in the mass matrix ($U_{e3}$) can be very small (here $\nu_3$ is the third mass eigenstate, that we assume to be mainly $\nu_\tau$). There is an upper bound on $U_{e3}$, coming from neutrinoless double-beta decay, that is is not difficult to accomodate. Unless the heavy $\nu_\tau$ is a Dirac neutrino (and, as explained below, it must be Majorana for most of the mass range of interest) it will contribute with amplitude $mU_{e3}^2$ to the effective $\nu_e$ Majorana mass measured in $\beta\beta_{0\nu}$ decay. The bound obtained, $U_{e3}^2 \lesssim (2\,{\rm eV}/m)$, (the $\beta\beta_{0\nu}$ lab limit is currently $m_{\rm eff} \lesssim 2$ eV) is more restrictive than the present experimental upper bound of about 0.0175 (for $\sin^2 2\theta \lesssim 0.07$ and $\Delta m^2 \gtrsim 10{\rm eV}^2$; [38]), but not difficult to obtain in models.

As mentioned above, constraints from energy loss due to the emission of inert right handed neutrinos in SN1987A rule out $\nu_\tau$ Dirac masses in the range $O(10{\rm keV})$ to 1MeV [39,40,41,42,43]. A related but independent bound from SN1987A [44] excludes lifetimes between $3.2 \times 10^{-17}(m/{\rm keV})$ yr and $1.6(m/{\rm keV})$ yr for Dirac masses in the range 1 keV to 300 keV. There is a more dubious bound that would also apply to Majorana neutrinos in the same mass range, if the dominant decay is $\nu_\tau \to \nu_e\phi$, which rules out lifetimes in the range $10^{-2}({\rm keV}/m)$ yr to $600({\rm keV}/m)$ yr [37,45]. Energy loss arguments applied to the almost inert boson $\phi$, only require the lifetime to be longer that the duration of the neutrino pulse seen from SN1987A, $\tau \gtrsim 10\,(m/E)$s $\simeq 10^{-3}(m/{\rm keV})$s [35], which becomes relevant only for the heavier $\nu_\tau$ proposed by [14]. Also in contrast to this heavier neutrino, our $\nu_\tau$ have no effect on nucleosynthesis [15,46,47], and are therefore easier to accommodate in particle physics models.

The decay mode into three neutrinos is more difficult to implement in phenomenologically viable models than the decay into a neutrino and a Goldstone boson. For the latter only singlet or mostly-singlet-mixed Majorons are allowed by the LEP bound on the number of equivalent neutrino species, essentially $N_\nu = 3$, measured in the invisible decay width of the $Z^0$ boson. These models naturally fit well the relatively long lifetimes required here. Typical non-minimal singlet Majoron models (and other models, such as familons or complicated leptonic symmetries) produce lifetimes of order $\tau \simeq V^2/m^3 = 2 \times 10^6 (V/10^{10}{\rm GeV})^2 (m/{\rm keV})^{-3}$yr, where $V$ is the spontaneous symmetry breaking scale (see for example [37] and references therein). The desired lifetimes of Fig. 4 correspond to very reasonable values of $V$ around $10^8 - 10^9$GeV. Moreover the non-standard annihilations predicted in these models are much smaller than weak (so that the relic density of neutrinos is that predicted by the Standard Model, as assumed in this paper). We speak of non-minimal singlet Majoron models because in the minimal singlet Majoron model [48], with just one singlet Higgs field, the decay rates are suppressed, $\tau \simeq V^4/m^5$ [49, 50]. This makes the required spontaneously broken scale $\mathcal{O}(10{\rm GeV})$ leading to phenomenologically more complicated models, if they are viable at all.

## 6. Conclusions

In models with decaying neutrinos the product of neutrino mass$^2$ and decay time ($m^2\tau$) is strongly constrained by large-scale structure models. The predicted matter and radiation power spectra are *approximately* degenerate in

$$h \left(1 + 0.1 \left[m_{\text{keV}}^2 \tau_{\text{yr}}\right]^{2/3}\right)^{-1/2} \quad (15)$$

when $\Omega_0 = 1$. We caution however that this formula is only approximate, and it is better to use the results of Table 1. Lowering $\Omega_0$ or $h$, or introducing decaying neutrinos, provide a shift in the epoch of matter–radiation equality, which improves the agreement of the predicted matter power spectrum with the large–scale structure data in CDM models. The latter two scenarios are distinguishable from the former by studying the positions of the peaks in the CMB radiation power spectrum on degree scales and smaller.

In the future we believe that $H_0$ will be measured at cosmological distances via gravitational lensing of variable quasars and the Sunyaev-Zel'dovich effect in galaxy clusters, as well as being determined locally by a plethora of mostly more indirect measurements. Thus one may be optimistic that, if neutrino decays occur in the relevant parameter range to resolve the large–scale structure problem in a CDM model, $m^2\tau$ can eventually be determined from the CMB. If this turns out to be the case then neutrino physics can be studied via temperature fluctuations in the CMB. At present one can already rule out a large range of the parameter space from large–scale structure data, see Figs. 2,4.

**Acknowledgements**

We would like to thank Douglas Scott and Wayne Hu for several useful conversations and comments on this and related work. This research was supported by the NSF and the TNRLC. G.G. was supported in part by the Department of Energy under Grant DE-FG03-91ER 40662 TaskC.

**Table**

| $m$ (keV) | $\tau$ (yr) | $\Gamma$ | $V_{60}$ | $V_{40}$ | $\sigma_8$ |
|---:|---:|---:|---:|---:|---:|
| 0.1 | 0.1 | 0.51 | 304. | 382. | 1.23 |
| 0.1 | 1.0 | 0.51 | 304. | 382. | 1.23 |
| 0.1 | 10.0 | 0.52 | 303. | 381. | 1.22 |
| 0.1 | 100.0 | 0.52 | 300. | 375. | 1.18 |
| 1.0 | 0.1 | 0.51 | 303. | 381. | 1.22 |
| 1.0 | 1.0 | 0.49 | 300. | 375. | 1.17 |
| 1.0 | 10.0 | 0.43 | 284. | 352. | 0.98 |
| 1.0 | 100.0 | 0.34 | 249. | 299. | 0.65 |
| 10.0 | 0.1 | 0.41 | 284. | 352. | 0.97 |
| 10.0 | 1.0 | 0.29 | 249. | 299. | 0.64 |
| 10.0 | 10.0 | 0.16 | 193. | 222. | 0.32 |
| 10.0 | 100.0 | 0.09 | 132. | 145. | 0.13 |
| 100.0 | 0.1 | 0.16 | 193. | 222. | 0.32 |
| 100.0 | 1.0 | 0.08 | 132. | 145. | 0.13 |
| 100.0 | 10.0 | 0.04 | 84. | 89. | 0.05 |
| 100.0 | 100.0 | 0.02 | 54. | 56. | 0.02 |

Table 1. Some measures of large scale power for cosmologies with decaying neutrinos, in the limit $\Omega_B \to 0$. We expect $\sim 10\%$ variation in these numbers for $\Omega_B = 1 - 10\%$. All models have been normalized to have CMB fluctuations consistent with the variance measured by COBE smoothed on $10°$. The shape parameter $\Gamma$ which best describes $T(k)$ is given along with the rms velocity (in km/s) in spheres of radius $60h^{-1}$Mpc and $40h^{-1}$Mpc after smoothing with a gaussian of $12h^{-1}$Mpc. In the last column we give the 'normalization' $\sigma_8$.

**Figures**

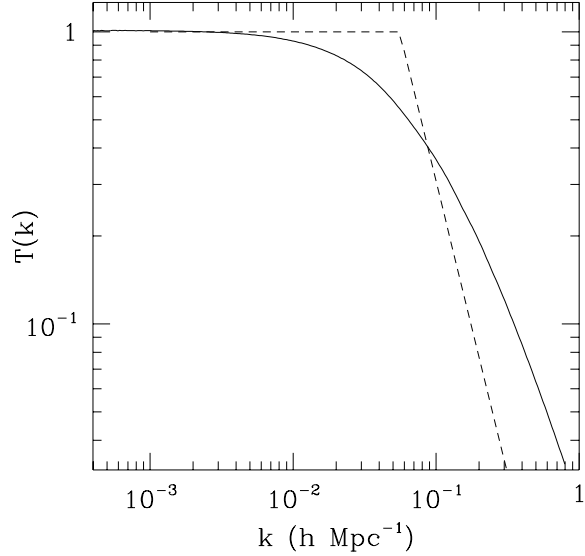

Fig. 1: The transfer function (solid) for a standard CDM model ($\Omega_0 = 1$, $\Omega_B = 5\%$, $h = 0.5$), along with the approximation (dashed) based on Eq.(6). Note that the approximation of a sudden "horizon crossing" is not quantitatively accurate. The transfer function is a factor of $\sim 3$ off at the $\sigma_8$ scale, indicating the need to accurately evolve the perturbations in order to use LSS to set limits on the parameter space.

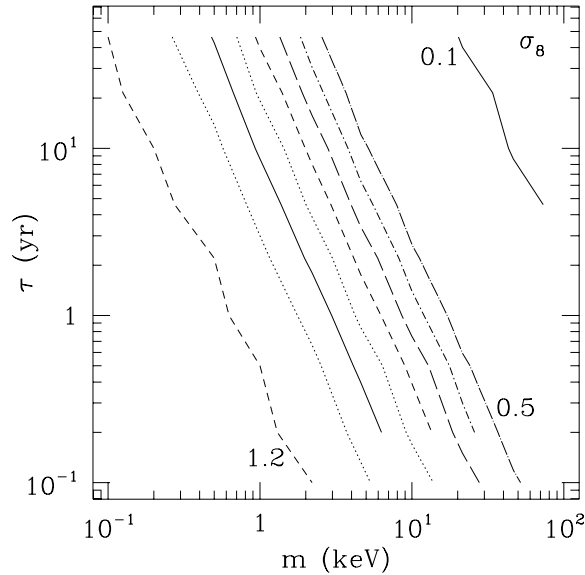

Fig. 2: A contour plot of $\sigma_8$ in ($m,\tau$) space. Contours are for $\sigma_8 = 0.5\ldots1.2$ in steps of 0.1, with $\sigma_8$ decreasing to the top right of the plot. The lower solid line is $\sigma_8 = 1.0$. We also show a (solid) contour with $\sigma_8 = 0.1$ (upper right).

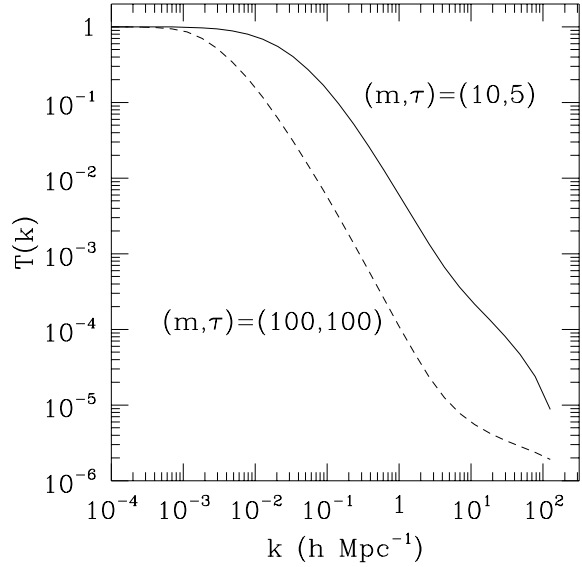

Fig. 3: The transfer functions for a model with $m = 10\,\text{keV}$ and $\tau = 5\,\text{yr}$ (solid) and for $m = 100\,\text{keV}$ and $\tau = 100\,\text{yr}$ (dashed). Notice that for the large mass and lifetime the first period of matter domination leads to a "plateau" in the transfer function at high $k$ (small scales). For the case $m = 10\,\text{keV}$ and $\tau = 5\,\text{yr}$, which has $\Gamma \simeq 0.25$ as preferred by LSS, the "plateau" is almost non-existent, though changes in the slope with $k$ are visible.

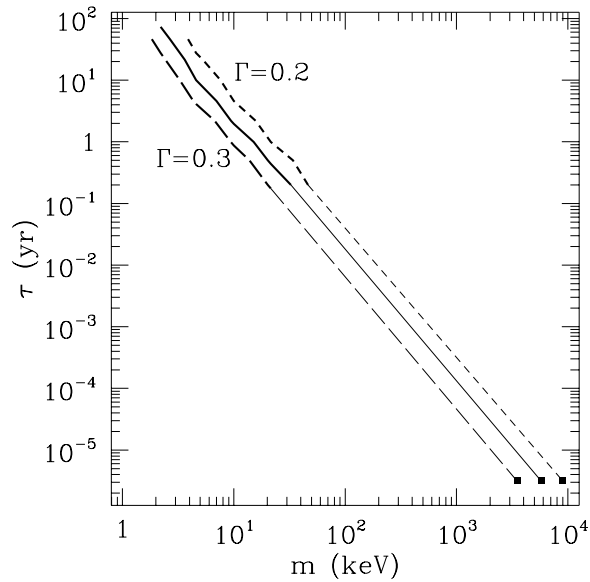

Fig. 4: A contour plot of $\Gamma$ in $(m,\tau)$ space which extrapolates to the larger masses and shorter lifetimes of the $\tau$CDM scenario [14]. The contours are for $\Gamma = 0.2, 0.25, 0.3$ (right to left). Heavy contours mark where we have specifically run models, while the lighter contours are the extrapolation to results quoted in [14] (the squares).

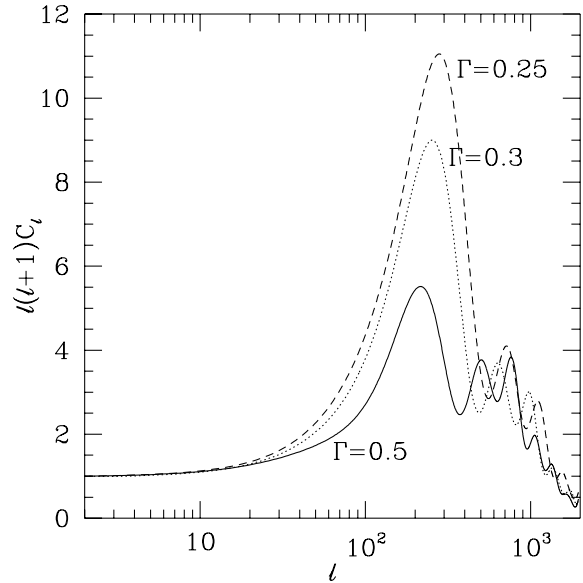

Fig. 5: The angular power spectrum of the CMB. The solid line is for standard CDM ($h = 0.5$, $\Omega_0 = 1$, $\Omega_B = 5\%$), the dashed line is for a model with $m = 10\,\mathrm{keV}$ and $\tau = 5\,\mathrm{yr}$ and the same $\Omega$ and $h$ as sCDM, which provides a good fit to the LSS data ($\Gamma \simeq 0.25$). Models with $\rho_\nu$ increased or $h$ lowered to impose $\Gamma = 0.25$ are almost identical. Also shown (dotted) is the power spectrum for $m = 6\,\mathrm{keV}$ and $\tau = 6\,\mathrm{yr}$ which has $\Gamma \simeq 0.3$. Note that the spectrum is quite sensitive to $\Gamma$.